\documentclass[twoside]{article}
\usepackage{amsmath,amssymb}
\input BoxedEPS
\SetRokickiEPSFSpecial  
\HideDisplacementBoxes 

\catcode`\@=11
\long\def\@makefntext#1{
\protect\noindent \hbox to 3.2pt {\hskip-.9pt
$^{{\eightrm\@thefnmark}}$\hfil}#1\hfill}               

\def\thefootnote{\fnsymbol{footnote}}
\def\@makefnmark{\hbox to 0pt{$^{\@thefnmark}$\hss}}    

\def\ps@myheadings{\let\@mkboth\@gobbletwo
\def\@oddhead{\hbox{}
\rightmark\hfil\eightrm\thepage}
\def\@oddfoot{}\def\@evenhead{\eightrm\thepage\hfil
\leftmark\hbox{}}\def\@evenfoot{}
\def\sectionmark##1{}\def\subsectionmark##1{}}

\oddsidemargin=\evensidemargin
\renewcommand{\thefootnote}{\fnsymbol{footnote}}

\newcounter{sectionc}\newcounter{subsectionc}\newcounter{subsubsectionc}
\renewcommand{\section}[1] {\vspace{12pt}\addtocounter{sectionc}{1}
\setcounter{subsectionc}{0}\setcounter{subsubsectionc}{0}\noindent
        {\tenbf\thesectionc. #1}\par\vspace{5pt}}
\renewcommand{\subsection}[1] {\vspace{12pt}\addtocounter{subsectionc}{1}
        \setcounter{subsubsectionc}{0}\noindent
        {\bf\thesectionc.\thesubsectionc. {\kern1pt \bfit #1}}\par\vspace{5pt}}
\renewcommand{\subsubsection}[1] {\vspace{12pt}\addtocounter{subsubsectionc}{1}
        \noindent{\tenrm\thesectionc.\thesubsectionc.\thesubsubsectionc.
        {\kern1pt \tenit #1}}\par\vspace{5pt}}
\newcommand{\nonumsection}[1] {\vspace{12pt}\noindent{\tenbf #1}
        \par\vspace{5pt}}

\newcounter{appendixc}
\newcounter{subappendixc}[appendixc]
\newcounter{subsubappendixc}[subappendixc]
\renewcommand{\thesubappendixc}{\Alph{appendixc}.\arabic{subappendixc}}
\renewcommand{\thesubsubappendixc}
        {\Alph{appendixc}.\arabic{subappendixc}.\arabic{subsubappendixc}}

\renewcommand{\appendix}[1] {\vspace{12pt}
        \refstepcounter{appendixc}
        \setcounter{figure}{0}
        \setcounter{table}{0}
        \setcounter{lemma}{0}
        \setcounter{theorem}{0}
        \setcounter{corollary}{0}
        \setcounter{definition}{0}
        \setcounter{equation}{0}
        \renewcommand{\thefigure}{\Alph{appendixc}.\arabic{figure}}
        \renewcommand{\thetable}{\Alph{appendixc}.\arabic{table}}
        \renewcommand{\theappendixc}{\Alph{appendixc}}
        \renewcommand{\thelemma}{\Alph{appendixc}.\arabic{lemma}}
        \renewcommand{\thetheorem}{\Alph{appendixc}.\arabic{theorem}}
        \renewcommand{\thedefinition}{\Alph{appendixc}.\arabic{definition}}
        \renewcommand{\thecorollary}{\Alph{appendixc}.\arabic{corollary}}
        \renewcommand{\theequation}{\Alph{appendixc}.\arabic{equation}}
        \noindent{\tenbf Appendix \theappendixc #1}\par\vspace{5pt}}
\newcommand{\subappendix}[1] {\vspace{12pt}
        \refstepcounter{subappendixc}
        \noindent{\bf Appendix \thesubappendixc. {\kern1pt \bfit #1}}
        \par\vspace{5pt}}
\newcommand{\subsubappendix}[1] {\vspace{12pt}
        \refstepcounter{subsubappendixc}
        \noindent{\rm Appendix \thesubsubappendixc. {\kern1pt \tenit #1}}
        \par\vspace{5pt}}

\newcommand{\textlineskip}{\baselineskip=13pt}
\newcommand{\smalllineskip}{\baselineskip=10pt}

\def\eightcirc{
\begin{picture}(0,0)
\put(4.4,1.8){\circle{6.5}}
\end{picture}}
\def\eightcopyright{\eightcirc\kern2.7pt\hbox{\eightrm c}}

\def\abstracts#1#2#3{{
        \centering{\begin{minipage}{4.5in}\baselineskip=10pt\footnotesize
        \parindent=0pt #1\par
        \parindent=15pt #2\par
        \parindent=15pt #3
        \end{minipage}}\par}}

\newcommand{\bibit}{\nineit}

\renewenvironment{thebibliography}[1]
        {\frenchspacing
         \ninerm\baselineskip=11pt
         \begin{list}{\arabic{enumi}.}
        {\usecounter{enumi}\setlength{\parsep}{0pt}
         \setlength{\leftmargin 12.7pt}{\rightmargin 0pt} 
         \setlength{\itemsep}{0pt} \settowidth
        {\labelwidth}{#1.}\sloppy}}{\end{list}}

\newcounter{itemlistc}
\newcounter{romanlistc}
\newcounter{alphlistc}
\newcounter{arabiclistc}

\newcommand{\fcaption}[1]{
        \refstepcounter{figure}
        \setbox\@tempboxa = \hbox{\footnotesize Fig.~\thefigure. #1}
        \ifdim \wd\@tempboxa > 5in
           {\begin{center}
        \parbox{5in}{\footnotesize\smalllineskip Fig.~\thefigure. #1}
            \end{center}}
        \else
             {\begin{center}
             {\footnotesize Fig.~\thefigure. #1}
              \end{center}}
        \fi}

\newcommand{\tcaption}[1]{
        \refstepcounter{table}
        \setbox\@tempboxa = \hbox{\footnotesize Table~\thetable. #1}
        \ifdim \wd\@tempboxa > 5in
           {\begin{center}
        \parbox{5in}{\footnotesize\smalllineskip Table~\thetable. #1}
            \end{center}}
        \else
             {\begin{center}
             {\footnotesize Table~\thetable. #1}
              \end{center}}
        \fi}

\def\@citex[#1]#2{\if@filesw\immediate\write\@auxout
        {\string\citation{#2}}\fi
\def\@citea{}\@cite{\@for\@citeb:=#2\do
        {\@citea\def\@citea{,}\@ifundefined
        {b@\@citeb}{{\bf ?}\@warning
        {Citation `\@citeb' on page \thepage \space undefined}}
        {\csname b@\@citeb\endcsname}}}{#1}}

\newif\if@cghi
\def\cite{\@cghitrue\@ifnextchar [{\@tempswatrue
        \@citex}{\@tempswafalse\@citex[]}}
\def\citelow{\@cghifalse\@ifnextchar [{\@tempswatrue
        \@citex}{\@tempswafalse\@citex[]}}
\def\@cite#1#2{{$\null^{#1}$\if@tempswa\typeout
        {IJCGA warning: optional citation argument
        ignored: `#2'} \fi}}

\def\pmb#1{\setbox0=\hbox{#1}
        \kern-.025em\copy0\kern-\wd0
        \kern.05em\copy0\kern-\wd0
        \kern-.025em\raise.0433em\box0}

\def\fnt#1#2{\footnotetext{\kern-.3em
        {$^{\mbox{\scriptsize #1}}$}{#2}}}

\def\fpage#1{\begingroup
\voffset=.3in
\thispagestyle{empty}\begin{table}[b]\centerline{\footnotesize #1}
        \end{table}\endgroup}

\def\runninghead#1#2{\pagestyle{myheadings}
\markboth{{\protect\footnotesize\it{\quad #1}}\hfill}
{\hfill{\protect\footnotesize\it{#2\quad}}}}
\font\tenrm=cmr10
\font\tenit=cmti10
\font\tenbf=cmbx10
\font\bfit=cmbxti10 at 10pt
\font\ninerm=cmr9
\font\nineit=cmti9

\font\eightrm=cmr8

\def\qed{\hbox{${\vcenter{\vbox{                        
   \hrule height 0.4pt\hbox{\vrule width 0.4pt height 6pt
   \kern5pt\vrule width 0.4pt}\hrule height 0.4pt}}}$}}

\renewcommand{\thefootnote}{\fnsymbol{footnote}}        

\newcommand{\fract}[2]{{\textstyle\frac{#1}{#2}}}
\begin{document}

\runninghead{Frank Wilczek} {Future Summary}

\normalsize\textlineskip
\thispagestyle{empty}
\setcounter{page}{1}

\centerline{\footnotesize  MIT-CTP\#3072 \hfill hep-ph/0101187}

\vspace*{0.125truein}

\fpage{1}
\centerline{\bf FUTURE SUMMARY\footnote{Closing talk
delivered at the LEPfest, CERN, October 11, 2000.  It closely resembles the
closing talk I delivered at the DPF meeting, Columbus, Ohio, August 2000.  
It will be published in the 
\emph{Proceedings}.}}
\vspace*{0.125truein}
\centerline{\footnotesize FRANK WILCZEK}
\vspace*{0.01truein}
\centerline{\footnotesize\it Center for Theoretical Physics, Massachusetts
Institute of Technology, 6-305}
\baselineskip=10pt
\centerline{\footnotesize\it Cambridge, MA 02139-4307,
USA}
\vspace*{0.125truein}

\vspace*{0.21truein}
\abstracts{We are emerging from a period of consolidation in particle
physics.  Its great, historic achievement was to establish the Theory of
Matter.  This Theory will serve as our description of ordinary matter under
ordinary conditions -- allowing for an extremely liberal definition of
``ordinary" -- for the foreseeable future.   Yet there are many indications,
ranging from the numerical to the semi-mystical, that a new fertile period lies
before us.  We will discover compelling evidence for the unification of
fundamental forces and for new quantum dimensions (low-energy
supersymmetry).  We will identify new forms of matter, which dominate the
mass density of the Universe.  We will achieve much better fundamental
understanding of the behavior of matter in extreme astrophysical and
cosmological environments.   Lying beyond these expectations, we can
identify deep questions that seem to call for ideas outside our present grasp.
And there's still plenty of room for surprises.}{}{}

\textlineskip                    
\vspace*{12pt}

\noindent
It is altogether appropriate, I think, to end this celebration of the
completion
of a great scientific program, and the retirement of a great machine, with a
look to the future.  The historic achievement of LEP has been to establish,
with an astonishing degree of rigor and beyond all reasonable doubt, what
will stand for the foreseeable future -- perhaps for all time -- as the working
Theory of Matter.

Many years ago, on the occasion of the founding of the Cavendish lab, James
Clerk Maxwell said
\begin{quote}
The history of science shows that even during that phase of her
progress in which she devotes herself to improving the accuracy of the
numerical measurements of quantities long familiar, she is preparing the
materials for the subjugation of new regions, which would have remained
unknown if she had been contented with the rough methods of her early
pioneers.
\end{quote}

These words of Maxwell have turned out to be prophetic, of course, many
times over.  Precision measurements of the blackbody spectrum helped lead
to early quantum theory.  Precision measurements on the spectrum of
hydrogen helped lead to modern quantum mechanics and ultimately to
quantum field theory.  Precision measurements on the K-meson system
helped lead, in one way or another, to the discovery of parity violation, CP
violation, and charm.  Precision measurements in deep inelastic scattering
helped lead to modern quantum chromodynamics, or QCD.   I believe, for
reasons I will enumerate shortly, that they apply again today.  For the
precision measurements made at LEP, besides establishing the Theory of
Matter, give us some very definite and specific clues for what lies beyond
that Theory.   They hint at new worlds of phenomena, connected with the
completion and structural unification of the Theory of Matter, and with the
existence of new quantum dimensions.   It is fitting, poetic -- and very
exciting -- that the systematic exploration of these new worlds will very
likely commence with the commissioning of LEP's literal descendant, the
Large Hadron Collider (LHC).

\textheight=7.8truein
\setcounter{footnote}{0}
\renewcommand{\thefootnote}{\alph{footnote}}
\bigskip

\section{The Theory of Matter}
\noindent Before discussing where we might be going it is reasonable to
reflect on where we are, and how we got there.

In 1900, physics had a completely different character from what it has
today.  It described {\it how \/} matter will evolve, given its initial condition.
Classical physics cannot explain why there are material substances with
definite, reproducible properties at all, much less why there are just the
particular molecules, atoms, and nuclei with the specific properties we
observe in Nature.    In short, classical physics cannot address questions about
{\it what \/} matter is, or {\it why} it is that way.

Modern physics does address these ``what" and ``why" questions.  Indeed,
most of us believe -- for very solid reasons, I think -- that we have in hand,
formulated quite precisely, the laws that in principle answer the central
``what" and questions about matter, and advance the ``why" questions to new
levels of abstraction and sophistication\footnote{In practice, of course, we can
only deduce quantitative consequences directly from the fundamental laws in
special, simple cases.}.   It is a great achievement, of historic proportions.

When the modern theories of matter, based on relativistic quantum field
theory, the gauge principle, spontaneous symmetry breaking, and asymptotic
freedom, were first proposed, they were provisional and hypothetical. Indeed,
for some years there were various competing ``models" for electroweak
interactions, there was considerable skepticism that straight local quantum
field theory is adequate to describe the strong interaction, and the
experimental evidence for both theories was meager.   In those
circumstances, it was appropriate to speak of a ``Standard Model" of particle
physics.  It seems to me however that by now this name no longer does
justice to what has been achieved.  Far more grandiose names have been used
for far less substantial achievements.  So I propose to call the ``Standard
Model" what we now know it to be -- the Theory of Matter.  More precisely, I
propose `Theory of Matter' to refer to the core concepts (quantum field
theory, gauge symmetry, spontaneous symmetry breaking, asymptotic
freedom) and the assignments of the lightest quarks and leptons.  These
concepts provide us with an extraordinarily powerful, economical description
of matter.  It will never erode, whatever is eventually discovered ``Beyond the
Standard Model".  We can keep `Standard Model' to describe the
(time-dependent!) minimalist position on more negotiable bits, such as the
number of Higgs doublets.

\subsection{Pre- and Post-LEP}
\noindent The Theory of Matter has two distinct, but smoothly meshing
components: the
$SU(2) \times U(1)$ electroweak gauge theory, and the $SU(3)$ color gauge
theory, QCD.   When LEP began operation, the frameworks for each of these
theories were in place, but neither had been nailed together very tightly, and
they could not support much weight.    Now it is very different.

\begin{figure}[ht]
\centerline{\BoxedEPSF{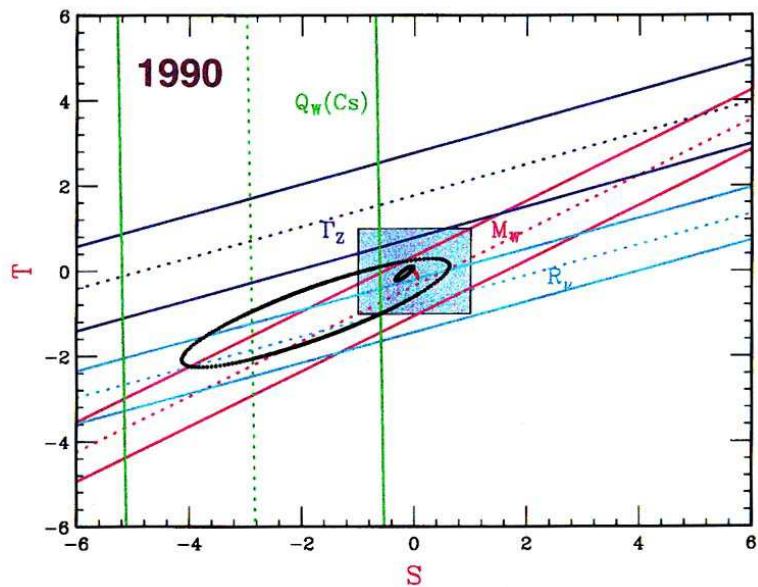 scaled 800}}
\caption{Data on fundamental electroweak parameters in 1990.    There was
broad consistency with minimal $SU(2)\times U(1)$, but  little sensitivity to
radiative corrections.}\label{f1}
\end{figure} Let's compare, for example, the state of the determination of
electroweak parameters in 1990 and today.  Figure~\ref{f1}, although it
represents only a small selection of the results, conveys their character.   In
1990, there was essentially no sensitivity to weak-scale radiative corrections,
and in particular no very meaningful constraints on the properties of
not-yet-observed components of the Standard Model ($t$ quark, Higgs boson)
from their virtual effects.

\begin{figure}[ht]
\centerline{\BoxedEPSF{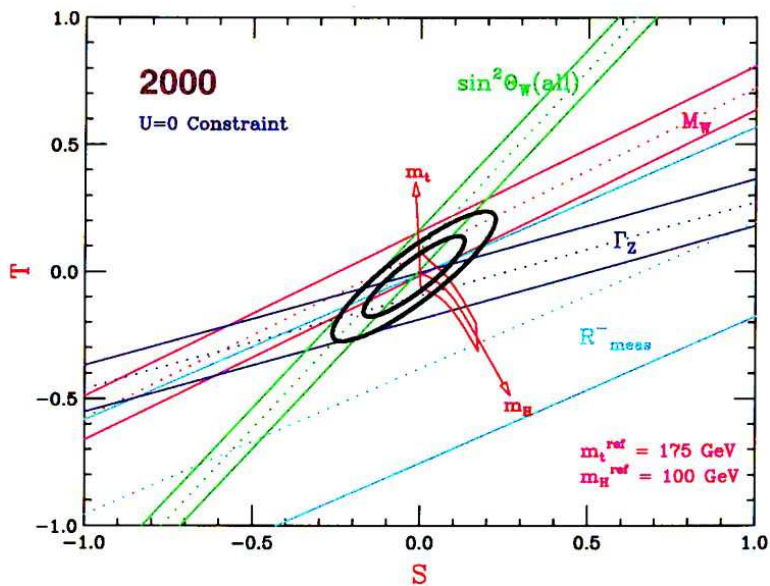 scaled 800}}
\caption{Data on fundamental electroweak parameters in 2000.   Careful
inclusion of the radiative corrections, including loops  containing both $W$
and $Z$ bosons and the color gluons of QCD,  is necessary to do justice to the
data.    One can discriminate the  effects of the top quark mass and the Higgs
boson mass.}\label{f2}
\end{figure}

Now, largely thanks to work at LEP, we have exquisite results for many
independent observables.  To visualize the difference, note that
Figure~\ref{f2}, displaying the 2000 data, is a blow-up of the tiny central
ellipse in Figure~\ref{f1}.  The newer results provide stringent consistency
tests for the Theory of Matter. For example, the agreement between
experiment and theory is adequate only after nonlinear gauge boson
interactions and multi-loop virtual gluon exchange are properly included. The
results are also accurate enough to resolve radiative corrections due to heavy
particles, Among other things, this work indicated the remarkable heaviness
of the $t$ quark before that particle was observed.

Looking to the future, these results provide powerful guidance in searching
for the Higgs particle, in assessing the plausibility of technicolor or
large-extra-dimension scenarios as compared with low-energy
supersymmetry, and in formulating ideas for unified field theories, as I shall
discuss below.

\begin{figure}[ht]
\centerline{\BoxedEPSF{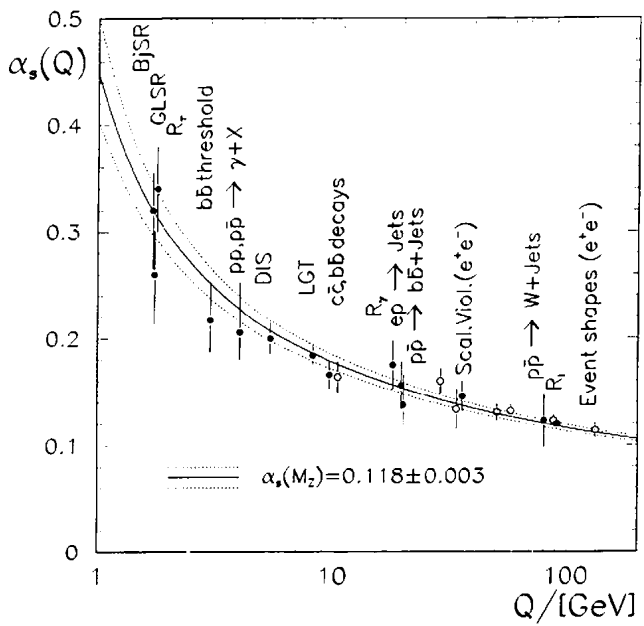 scaled 1000}}
\caption{Data from an enormous number and variety of  experiments, all
accurately described by QCD, and together  demonstrating the predicted
shrinkage of the strong coupling at high  energy (asymptotic
freedom).}\label{f3}
\end{figure}

Turning to QCD, the classic ``picture worth a thousand words" is
Figure~\ref{f3}.  It shows the running of the strong coupling.  The LEP points
mostly appear at the high-energy end.  Of course, summarizing the hundreds
of QCD tests done at LEP in a couple of points hardly does them justice.  There
is a very rich and extensive story here, with chapters including direct tests of
flavor-universality, studies of color coherence, and beautiful work, leading to
another determination of $\alpha_s$, on $\tau$ decays.

But time is limited, so I will mention just one especially remarkable aspect of
the experimental tests of QCD.   As you can see from Figure~\ref{f3}, the QCD
prediction for the running of couplings has a focussing property: a fairly wide
range of values at low energy scales, or equivalently of the scale parameter
$\Lambda_{\rm QCD}$, implies accurately the same value for the strong
coupling $\alpha_s(M_W)$ governing LEP results.  Also, of course, the precise
values of quark mass parameters become irrelevant at these high energies.
Thus the QCD predictions for LEP represent essentially {\it zero parameter \/}
predictions for a host of measurable quantities, such as relative frequency of
two-, three- and four-jet events, angular and energy distributions, and the
variation of all these with energy.  There's no wiggle room.  If the Theory is
right, all these predictions had better hold true -- and they do.

The deepest-going results are sometimes those which become so embedded in
our world-view that we take them for granted.   Along this line, the Theory of
Matter has made it credible, in a way that as recently as 30 years ago it was
not, that we can succeed in understanding, quantitatively and in detail, the
``what" as well as the ``how" of Nature.  No amount of faith or philosophy is as
convincing as a few dozen successful two-loop calculations!

\section{Completion? (The Higgs Particle)}
\begin{figure}[ht]
\centerline{\BoxedEPSF{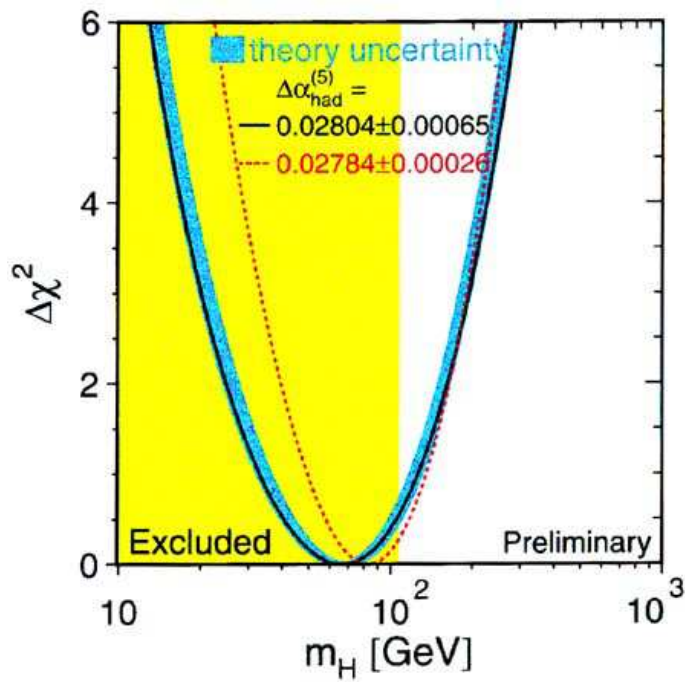 scaled 1200}}
\caption{Experimental constraints on the mass of the Higgs boson,  derived
from the study of radiative corrections, interpreted in the  framework of the
minimal Standard Model.}\label{f4}
\end{figure}

\noindent The Higgs particle is the only ingredient of the Standard Model that
has not yet been observed directly.   From the study of radiative corrections to
electroweak parameters, as indicated in Figure~\ref{f2}, one can infer limits on
the Higgs particle mass.  These limits assume, of course, that no additional
unknown particles are contributing.   They are displayed in a more expansive
format in Figure~\ref{f4}.

\begin{figure}[ht]
\centerline{\BoxedEPSF{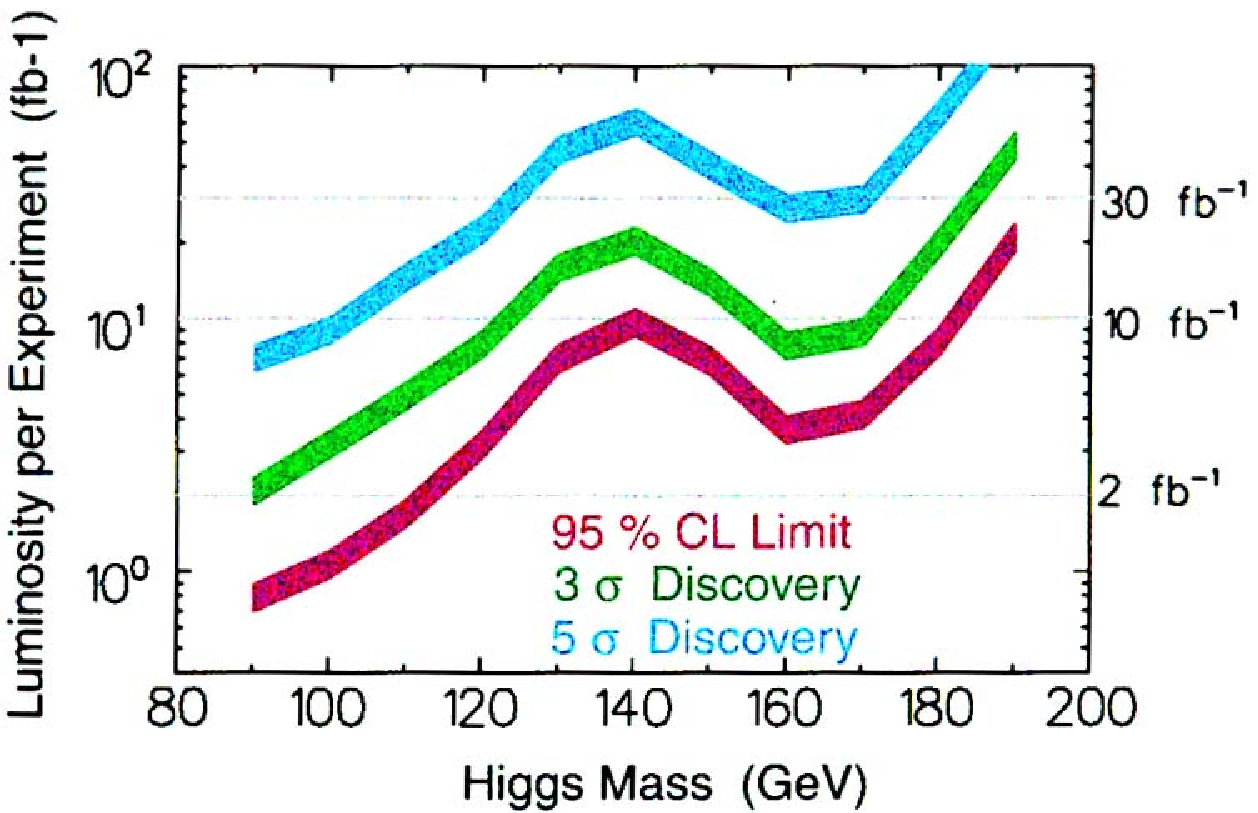 scaled 800}}
\caption{Prospects for discovery of the Higgs boson at the  Tevatron, as a
function of its mass and the accumulated luminosity.}\label{f5}
\end{figure}

It is quite impressive and significant how well the mass is boxed in.   This
makes the challenge facing the Fermilab Tevatron very tangible and
concrete.   That challenge is made graphically evident in Figure~\ref{f5}. I'd
like to advertise the interesting, though controversial, possibility of looking
for the Higgs signal in $\bar b b$ events with rapidity gaps, which could
further improve the prospects\cite{ref1}.  More theoretical work is needed
here.

\subsection{Some Sense of Proportion}
\noindent There is no doubt that discovery of the Higgs particle would be a
wonderful event.  However I do think it is important to keep a realistic sense
of proportion, and to be clear just what would be wonderful about it, and
what it would mean.

First of all, what is important is not so much the existence of a new highly
unstable particle -- there are lots of such particles, after all -- but the idea it
embodies: the idea of spontaneous symmetry breaking. An analogy: there is a
useful and intellectually rich theory of spontaneous chiral symmetry breaking
in QCD, but the analogue of the `Higgs' -- namely, the $\sigma$ meson -- is
hardly crucial to that theory; indeed, its nature and even its existence is still
debated.

From a conceptual perspective, the essential thing is not so much the Higgs
particle, but the doublet of which it is a member, and the dynamics it
implements.  And out of that (complex) doublet, three out of four components
have already been discovered, and studied in great detail, doing their
dynamical duty!  I mean, of course, the longitudinal components of the
$W^{\pm}$ and $Z$.

Second, the Higgs particle (or the doublet)  is certainly {\it not} -- despite
much loose talk to the contrary -- the Origin of Mass.   (Still less is it the God
Particle, whatever that means.)  Most of the mass of ordinary matter is
concentrated in protons and neutrons.  It arises from an entirely different,
and I think more profound and beautiful, source. Numerical simulation of QCD
shows that if we built protons and neutrons in an imaginary world with no
Higgs mechanism -- purely out of quarks and gluons with zero mass -- their
masses would not be very different from what they actually are.  Their mass
mostly arises from pure energy, associated with the dynamics of confinement
in QCD, according to relation $m = E/c^2$.   This profound account of the origin
of mass is a crown jewel in our  Theory of Matter.

Third, the Higgs particle is most unlikely to be an isolated phenomenon.  More
likely, it is the tip of an iceberg.  As I'll document shortly, there are good
reasons to believe that there are at least two complex doublets scalar ``Higgs''
doublets -- and thus at least five real particles, plus the three appearing as
longitudinal vector mesons.

So to me, the most exciting aspect of the discovery of a Higgs particle (or
particles) will be that the value of its mass will provide concrete guidance
regarding extension of the Theory of Matter.   For example, my favorite
extension -- minimal supersymmetry, effectively broken at nearly
electroweak energies, but with a clear separation of scales -- predicts a
relatively light Higgs particle.    The upper bounds extend to 130 Gev or so,
but they are not easy to saturate, and I'd be happier with 115 Gev.   On the
other hand, if we had the minimal Standard Model, a Higgs particle of this
mass would be cause for concern, since it would indicate that we live in a
metastable vacuum!

\section{Structure?}
\noindent Because the Theory of Matter is so successful, we should hold it to
high standards, and take its shortcomings very seriously.   Perhaps the most
profound of these shortcomings, because they relates so closely to the core
concepts, leap out from Figure~\ref{f6}.
\bigbreak

\begin{figure}[ht]
$$\begin{tabular}{ccc}
$\begin{pmatrix} u & u & u\\ d & d & d 
\end{pmatrix}_{1/6}$ & $\bigl( u^c \, u^c\, u^c\bigr)_{-2/3}$ & $\bigl( d^c \,
d^c\, d^c\bigr)_{1/3}$\\
 & $\begin{pmatrix}
\nu\\
 e 
\end{pmatrix}_{-1/2} \, e^c_1$ 
\end{tabular}
$$
\centerline{\hbox to 3truein{\hrulefill}}
\smallskip
$$
\begin{tabular}{lccccc}
    & R & W & B & G & P\\
$u$   & + & $-$ & $-$ & + & $-$ \\
$u$   & $-$ & + & $-$ & + & $-$ \\
$u$   & $-$ & $-$ & + & + & $-$ \\
$d$   & + & $-$ & $-$ & $-$ & + \\
$d$   & $-$ & + & $-$ & $-$ & + \\
$d$   & $-$ & $-$ & + & $-$ & + \\
$u^c$ & $-$ & + & + & $-$ & $-$ \\
$u^c$ & + & $-$ & + & $-$ & $-$ \\
$u^c$ & + & + & $-$ & $-$ & $-$ \\
$d^c$ & $-$ & + & + & + & + \\
$d^c$ & + & $-$ & + & + & + \\
$d^c$ & + & + & $-$ & + & + \\
$\nu$ & + & + & + & + & $-$ \\
$e$ & + & + & + & $-$ & + \\
$e^c$ & $-$ & $-$ & $-$ & + & + \\ {\bf N} & $-$ & $-$ & $-$ & $-$ & $-$ \\
\end{tabular}\quad \begin{aligned} Y &= -\fract16 (\mathrm{R+W+B})\\
   &\qquad {}+\fract14 (\mathrm{G+P})
\end{aligned}$$
\caption{Top part: the organization of fermions in the lightest family,  based
on $SU(3)\times SU(2)\times U(1)$.   Bottom part:  Organization of the
fermions in the lightest family, based on the  spinor {\bf 16} representation
of $SO(10)$.}\label{f6}
\end{figure}

In upper part of this figure I have displayed the transformation properties of
the lightest quarks and leptons under the gauge groups $SU(3)\times
SU(2)\times U(1)$.   Left-handed fields are used exclusively, so we employ
charge conjugation $u^c$ to get the right-handed $u$ quark into the game,
through its (left-handed) conjugate.    $SU(3)$ acts horizontally, $SU(2)$ acts
vertically, and the hypercharge $U(1)$ assignments are indicated by
subscripts.

There are two evident shortcomings to this structure.  First, the particles fall
into five disconnected pieces.  Second, there is no evident rhyme or reason to
the hypercharge assignments.  They are simply chosen to fit experiment.

Along the same lines, the gauge symmetry falls apart into three independent
pieces.

All these shortcomings can be overcome, in a way I find quite pretty and
compelling, by building upon the core concepts of the Theory of Matter itself.

\subsection{Unification of Multiplets}
\noindent Escalating the concept of spontaneous symmetry breaking, it is
natural to ask whether the $SU(3)\times SU(2)\times U(1)$ of the Theory of
Matter, which breaks to $SU(3)\times U(1)$, might itself arise from breaking
of a larger symmetry.

As is by now well-known, the $SU(3)\times SU(2)\times U(1)$, and the
fermions fit snugly into an $SU(5)$.  Using a simple breaking scheme
(condensate in the adjoint {\bf 24} representation), and starting with
fermions in the antisymmetric tensor $\overline{\bf 10}$ and vector {\bf 5}
representations, we arrive at precisely the gauge groups and fermion
multiplets of the Theory of Matter, including the hypercharge assignments.
This is a highly non-trivial coincidence.  Since it cuts the number of multiplets
down from five to two, and uniquely fixing the hypercharge assignments, this
unification achieves substantial esthetic gains over its starting point.

Still more beautiful is the possibility of unification afforded by the slightly
larger group $SO(10)$.  Now the fermions all fit into a single spinor {\bf 16}
representation.  This is a particularly elegant representation, with remarkable
properties, as indicated in the bottom part of Figure~\ref{f6}.  The components
of the spinor representation can be specified by their transformation
properties under the diagonal $SO(2)\times SO(2)\times SO(2)\times
SO(2)\times SO(2)$.  These have the physical interpretation of values of five
color charges.   All possible combinations of charges $\pm \frac12$ are
allowed, subject to the constraint that the number of $+\frac12$ charges is
even.  From these abstract mathematical rules, the gauge multiplets of the
Theory of Matter arise, with the pattern observed in Nature.   In particular,
the hypercharges are uniquely predicted from the strong and weak charges,
according to the simple formula
$$ Y~=~ -\frac16 (R+W+B) +\frac14 (G+P)~.
$$

The spinor {\bf 16} contains, in addition to the fermions of the Theory of
Matter, an additional  particle $N$.  Since $N$ is a singlet under $SU(3)\times
SU(2)\times U(1)$, it has none of the standard interactions with matter, so its
``non-discovery'' does not pose immediate problems.  Indeed it plays a major
{\it constructive \/}  role in the theory of neutrino masses, as I shall discuss a
little later.

\subsection{Unification of Couplings}
\noindent Unified gauge symmetry requires universal gauge coupling
strength.  This does not hold, of course, in the Theory of Matter.    The $SU(3)$
coupling is observed to be larger than the $SU(2)$ coupling, which in turn is
larger than the $U(1)$ coupling.

Fortunately, as we have seen in Figure~\ref{f3}, a great lesson of the Theory
of Matter is that coupling constants evolve with energy.  The same sorts of
calculations that give us asymptotic freedom in the strong interaction allow us
to evolve, theoretically, the effective couplings up to large energy, or
equivalently short distance, scales.   If the Theory of Matter derives from a
larger gauge symmetry, spontaneously broken at a unique large energy scale,
we should expect that these couplings meet at a point.  Indeed, in running
from high to low energies, the couplings only started to diverge once the big
symmetry was broken.

\begin{figure}[ht]
\centerline{\BoxedEPSF{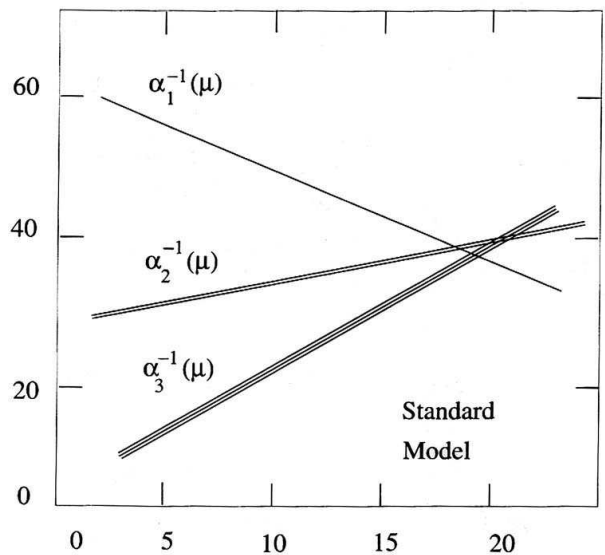 scaled 625}}
\caption{Near-unification of couplings, based on extrapolating the  running of
couplings in the minimal Standard Model.}\label{f7}
\end{figure}

If we evolve the couplings up to high energy using only the particles of the
Standard Model, we get the result shown in Figure~\ref{f7}.  Notice that to a
good approximation the inverse couplings are predicted to run
logarithmically, so the running generates straight lines in this log plot. The
width of the lines indicates the experimental uncertainties, post LEP.  It is a
remarkable near-miss; but a miss nonetheless.

One might, and many still do, try to repair this small discrepancy in any
number of ways, with slight perturbations on the Standard Model.   In the
absence of any powerful guiding principle, however, such fixes lack conviction.

\begin{figure}[ht]
\centerline{\BoxedEPSF{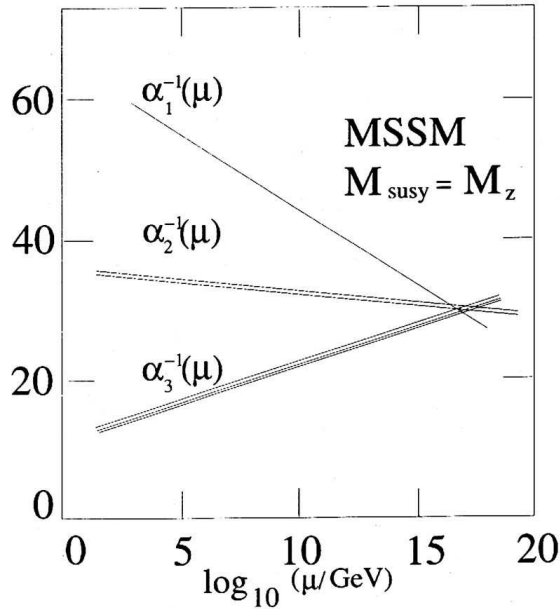 scaled 550}}
\caption{Accurate unification of couplings, based on extrapolating  the
running of couplings including the effects of a minimal  realization of
low-energy supersymmetry.}\label{f8}
\end{figure}

Much more compelling, I think, is to start with a deep idea, and to discover
that it unexpectedly solves a problem it wasn't originally specifically built
for.  Rather than tweaking the Standard Model, let us consider the apparently
drastic, but independently motivated, idea that supersymmetry is broken
only at relatively low ($\lesssim$ Tev) energies.   This modifies the running
of the couplings, in a way that is easy to compute, because there are more
virtual particles to consider.   If we extend the Standard Model in the most
economical way to include low-energy supersymmetry, we find the result
shown in Figure~\ref{f8}.      The unification now works extraordinarily
well.    This is a greatly encouraging result, both for unification and for
low-energy supersymmetry.

It was very surprising, at first, to discover that such a drastic modification of
the Standard Model caused only small changes in the predicted relation
among low-energy couplings.   After a bit of thought, however, it's not hard to
see why that relation is robust against certain classes of perturbations.
Basically, any addition of particles forming complete $SU(5)$ multiplets,
modestly split, will cause only small changes.   Indeed, the major reason that
the minimal supersymmetric result differs from the Standard Model result is
that low-energy supersymmetry requires two Higgs doublets which do not
have accompanying triplets, together of course with their fermionic
superpartners. (The triplets have very exotic quantum numbers and
potentially mediate rapid proton decay.  They must be super-heavy.)

\subsection{Significance}
\noindent The quantitative success of the unification of couplings calculation
(with low-energy supersymmetry) is undeniable.  What is its significance?

At the most formal level, the unification of couplings is an over-constrained
fit of three measured quantities -- $\alpha_1(M_{\rm W}), \alpha_2(M_{\rm
W}), \alpha_3(M_{\rm W})$ -- to two theoretical parameters, the scale of
unification and the strength of coupling at unification.  Given the precision of
the measurements, it is remarkable that a fit can be obtained.

But simply saying that one number falls into place does not do justice to the
state of affairs.   For there are many other things that could have gone wrong,
besides failure to find a good numerical fit.  If the couplings had met at too
small an energy scale, we would have difficulties with rapid proton decay.  If
they had met at a significantly larger a mass scale, at or above the Planck
scale, we would have had to worry about quantum gravity corrections.   The
actual scale at which they meet, not far on a logarithmic scale, but still
significantly,  below the Planck scale, is uniquely acceptable.  Similarly, if the
unified coupling were much larger we could not trust the perturbative
calculation.

\begin{figure}[ht]
\centerline{\BoxedEPSF{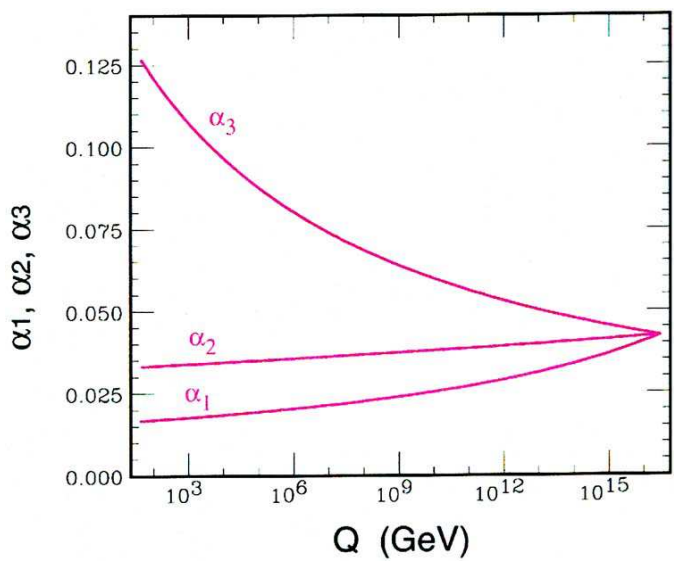 scaled 700}}
\caption{The same as Figure 8, using different variables.}\label{f9}
\end{figure}

I have heard it said, in reference to Figures 7 and 8, that ``two straight lines
will always meet in a point, and it's not so remarkable that three happen to''.
This attitude, I believe, is profoundly wrong-headed.   Some of my reasons
are those given in the previous paragraph.  Another is that the ``straight line''
nature of the running is in itself a profound result, reflecting the nature of
vacuum polarization in quantum field theory.  It appears semi-trivial only
because of the way it is plotted (inverse couplings on a log scale).   Perhaps
Figure~\ref{f9} is more impressive!

To my perception, the unification of multiplets and the unification of
couplings are the crown jewels of physics beyond the Standard Model.
Together, they make a powerful {\it prima facie \/} case for the elements that
went into their derivation: unified gauge symmetry, for the unification of
multiplets; renormalizable quantum field theory, operating smoothly up to
near-Planckian scales, for the proper logarithmic running of couplings; and
low-energy supersymmetry, for detailed numerical success.

Nowadays, in the context of string theory, we know -- or, rather, we have
incomplete suggestions about -- many alternative ways that the low-energy
$SU(3)\times SU(2)\times U(1)$ symmetry of the Theory of Matter might
emerge, from constructions that involve  neither effective unified gauge field
theories nor symmetry breaking through condensates.    Of course, there is no
necessary contradiction, since early reduction to an effective unified gauge
field theory also still remains a viable option.  Along this line, perhaps we
should take the striking apparent successes of the ``good old'' ideas I just
reviewed (and there are more to come!) as indications that in searching for
string-based models of Nature, we should look to those that reduce to
something like an effective supersymmetric $SO(10)$ renormalizable gauge
field theory just below the Planck scale.   Certainly, any other scheme has
some coincidences to explain.

\section{Seven Pillars of  Unification Wisdom}

I've just now discussed the first two of these:
\vspace*{-0.5pc}

\subsection{Multiplet Unification}
\vspace*{-1pc}

\subsection{Coupling Unification}

\noindent These unifications are motivated by the structure of the Theory of
Matter. They are firmly based on extrapolating the deep core concepts of that
Theory (quantum field theory, gauge symmetry, spontaneous symmetry
breaking, asymptotic freedom) to new energy scales.  They lead us into a
framework including unified gauge symmetry, renormalizable field theories
effective up to near-Planckian scales, and low-energy supersymmetry.   

This framework is usefully specific, and has several other desirable
consequences, that I'll summarize briefly now.

\subsection{Neutrino Mass Scale}

\noindent The oscillation of atmospheric neutrinos discovered by the SuperK
collaboration can be interpreted as evidence for a mass $m_{\nu_\tau}$ of
the tau neutrino $\nu_\tau$ of order $10^{-2}$ eV.   Within the framework of
electroweak $SU(2)\times U(1)$, this can be accommodated by means of a
non-renormalizable interaction
$$
\Delta {\cal L} ~=~ \frac1M \phi^\dagger l  \phi^\dagger l~,
$$ where $l$ is the lepton doublet and $\phi$ the Higgs doublet.  With $\phi$
replaced by its vacuum expectation value $v$, this becomes a Majorana
neutrino mass of magnitude
$v^2/M$.   With $M$ of order $10^{15}-10^{16}$ Gev, this is about right.  That
mass scale is equal to that which appears in the unification of couplings, as the
scale at which unification symmetry breaks.

There is a simple, concrete dynamical mechanism for generating neutrino
masses that explains this coincidence.   It involves the $N$ particle we met
before as the missing component of the $SO(10)$ spinor {\bf 16}.  Since it is
an $SU(3)\times SU(2) \times U(1)$ singlet, this particle can acquire a large
mass
$\sim M_{\rm U}$ at the scale where $SO(10)$ symmetry is broken, without
breaking those low-energy symmetries.  It also can connect to the
conventional left-handed neutrino $\nu$ by a normal Higgs-type mass term
acquiring a mass $m$.  By second-order perturbation theory, passing through
the intermediate $N$, we generate a Majorana mass of order $m^2/M_{\rm
U}$ for $\nu$.   Finally, we expect $m\sim v$ for the heaviest neutrino, since
this mass is related by symmetry to the large top quark mass (see below).

There are significant uncertainties in both steps of the argument, so this
calculation of the scale of neutrino masses is semi-quantitative at best. Still, it
is very impressive how the outlandishly small value of the neutrino mass,
relative to other quark and charged lepton masses, gets mapped to the
outlandishly large value of the unification scale, and how the existence of
$N$, at first sight an embarrassment, turns out to be a blessing.

\subsection{Basic Features of Electroweak Symmetry Breaking}
\noindent The point of departure for many ideas about physics beyond the
Standard Model is dissatisfaction with the minimal Standard Model account of
electroweak symmetry breaking.

In the minimal model electroweak symmetry breaking is, of course,
implemented by minimization of a simple potential for the Higgs doublet.  At
the classical level, and if we confine our gaze to the electroweak sector alone,
this would seem to be unobjectionable, and indeed very much in line with
how Occam might suggest that we parametrize our ignorance of the symmetry
breaking dynamics.   But if we consider the quantum version of the model, we
find ourselves in the somewhat distasteful situation of having quadratically
divergent radiative corrections to the Higgs doublet mass parameter.

That is not quite a contradiction, but it does beg the question of what
provides the cutoff.   The Standard Model by itself is not a well-behaved
quantum field theory.  It is not asymptotically free, and therefore most likely
it does not exist, nonperturbatively.  (Perturbation theory, which alone makes
the renormalization program plausible, goes bad in the ultraviolet.)   One
might reasonably expect that whatever additional physics we must add to the
Standard Model to make it a good theory is characterized by some much
larger  mass scale, simply because we've seen no direct sign of that physics. 
And then we have to understand why this larger mass scale does not infect
the Higgs doublet mass parameter, through radiative corrections.

The difficulty is exacerbated considerably if we take the unification of
couplings calculation seriously (as, of course, we should).   For this indicates a
unification scale of order $10^{16}$ Gev --  a whopping factor $10^{28}$ larger
than the electroweak scale, using the appropriate quadratic measure.  Any
corrections to the Higgs doublet mass arising from this sector must by highly
suppressed compared to naive dimensional analysis.   There are generally
both classical and quantum corrections.

Low-energy supersymmetry cleanly suppresses the quantum corrections, by
canceling off contributions between `nearly degenerate' ($\Delta M \lesssim$
Tev) virtual bosons and fermions.   Requiring adequate suppression gives a
condition of the rough form
$$ \frac\alpha\pi \Delta M^2 \lesssim v^2~,
$$ with $v$ as before.  This is the foundational argument, independent of the
unification of couplings calculation, for {\it low-energy \/} supersymmetry. It
is possible, and important, to be more precise about this naturality condition,
as we'll see shortly.

The suppression of classical corrections is a different, and much murkier,
question.  In fact it raises several issues: doublet-triplet splitting, the
so-called $\mu$ problem, the magnitude of soft supersymmetry breaking
terms, and perhaps others.   No simple or uniquely compelling answers are
available at present, so I'll say no more here.

This simple but powerful argument for low-energy supersymmetry could be,
and was, made before LEP.  During the LEP era, additional supporting
evidence has emerged.

The first, and most profound, piece of evidence may seem a little paradoxical:
it is how well the Standard Model has stood up to detailed quantitative
scrutiny.   To put this in perspective, we should contrast the approach of
low-energy supersymmetry with other attempts to address the problem of
stabilizing the weak scale.   

Instead of invoking cancellations to keep radiative corrections to the Higgs
mass small, one might imagine that there are form-factors.  For this, the Higgs
doublet must be composite on the weak scale, with some strong-coupling
dynamics to bind it.   That is the central idea of `technicolor' theories.   In
such theories, since there is no small coupling nor super-large mass
suppressing the new strong-coupling dynamics, there is no reason to expect
radiative corrections in general to be small.  One would expect, generically,
relatively large deviations from Standard Model predictions at the one-loop
level.  The situation deteriorates further if one tries to account for flavor
physics along these lines, since there are severe empirical bounds on the
expected neutral flavor-changing interactions.   

It is still worse if one tries to put unification or string physics at the weak
scale, since this would appear to bring in proton decay too.

\begin{figure}[ht]
\centerline{\BoxedEPSF{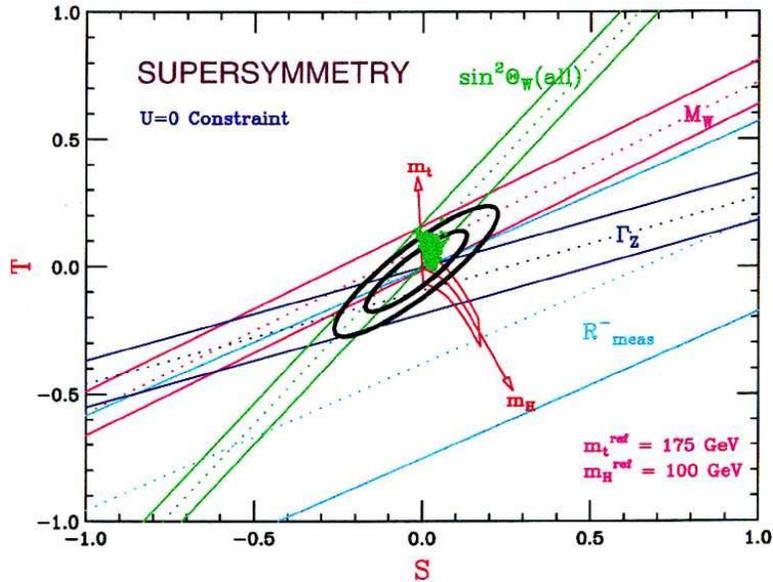 scaled 800}}
\caption{Radiative corrections to precision electroweak data as  calculated in
a minimal model of low-energy supersymmetry,  allowing for variation of the
unknown parameters of the model.}\label{f10}
\end{figure}

Perhaps some clever pastiche of tricks allows Nature to circumvent these
pitfalls\footnote{I am quite skeptical of this.  In particular, the idea that
proton decay can be suppressed by putting quarks and leptons on different
walls seems quite dubious to me.  Since the Standard Model itself supports
proton decay, albeit at imperceptible rates, through weak instantons, there
cannot be a universal suppression mechanism consistent with obtaining the
Standard Model as a low-energy limit.}.   But I prefer to think that 
Figure~\ref{f10} is pointing the way.  It shows how minimal implementations
of low-energy supersymmetry are easily consistent with precision
electroweak data.  Weak coupling and good ultraviolet behavior make this
self-effacement possible; the facts make it make mandatory.

A second piece of evidence is also visible in Figure~\ref{f10}.  Whereas
without supersymmetry the Higgs particle mass is essentially unconstrained,
supersymmetric models relate it to directly to the $Z$ mass, and after careful
calculation one finds masses below
$m_H \lesssim 130$ Gev in generic models.   This fits quite well with the
indirect observations, summarized in Figure~\ref{f4} -- and still better with
the possible discovery at $m_H \approx 115$ Gev.

\begin{figure}[ht]
\centerline{\BoxedEPSF{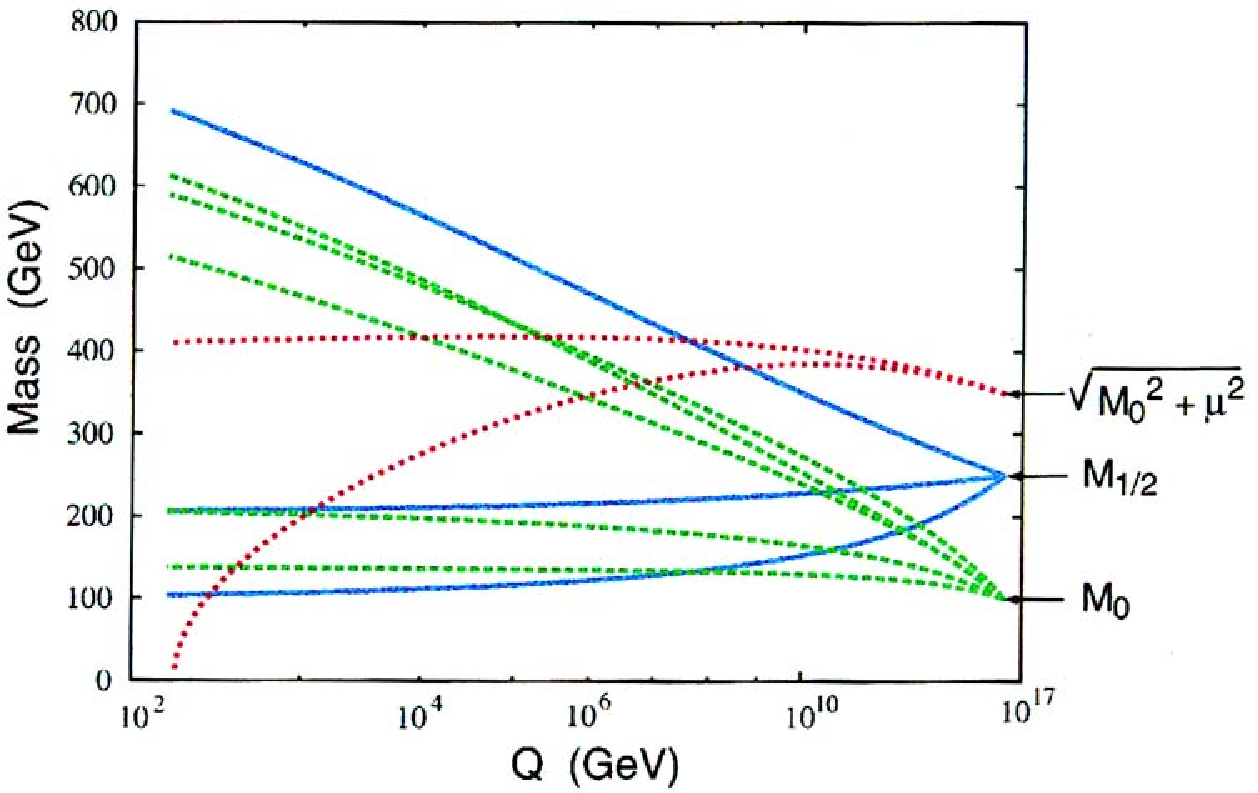 scaled 900}}
\caption{Running of effective masses down from the unification  scale, in a
minimal supergravity model.   The running is mainly  driven by radiative
corrections from virtual tops and stops.    It drives  the Higgs (mass)$^2$
parameter negative, inducing electroweak  symmetry breaking.}\label{f11}
\end{figure}

Finally, the observed large value of the top quark mass supports an elegant
mechanism for electroweak symmetry breaking, as shown in Figure~\ref{f11},
through running of the effective Higgs mass.   This nicely fills the
requirements of minimal supergravity models, with soft supersymmetry
breaking terms.

\subsection{Dark Matter Candidates}
 \noindent If baryon and lepton number are accurate symmetries, then so is
R-parity
$$ R ~=~ (-1)^{3B+L+2S}~,
$$ where $B$ is baryon number, $L$ is lepton number, and $S$ is spin.
Ordinary particles are even under R-parity; their supersymmetric partners
are odd.   Therefore the lightest supersymmetric particle is likely to be
extremely stable.   In many models of low-energy supersymmetry it is a
neutral fermion, generically called the neutralino.  The neutralino interacts
quite weakly with ordinary matter. Much study has been devoted to minimal
models of low-energy supersymmetry with universal soft breaking terms.  In
this framework, one generally finds that the neutralino is a linear
combination of bino and Higgsino (partners of the hypercharge gauge boson
and the Higgs particle). Given a concrete model, one can calculate the
production of neutralinos in the early Universe.  For a significant range of
parameters the calculated relic abundance of neutralinos, and their feeble
interactions with ordinary matter, makes them excellent candidates to supply
the missing mass that astronomers need.  I will discuss some specific
parameters and search strategies in a few moments.

A possible complication is that in plausible extensions of the minimal
framework, the lightest supersymmetric particle could have quite a different
character.  The neutralino (defined as the lightest R-odd member of the
Supersymmetric Standard Model) might well decay very slowly on particle
physics or laboratory time scales, but rapidly on cosmological time scales,
through ultra-weak interactions, into a still lighter R-odd particle.   A prime
candidate is the axino.   I find the idea that the missing mass is dominated by
axions, with quasi-stable neutralinos having decayed into axinos, quite
entertaining\cite{ref2}.

\subsection{Fermion Coupling Unification}
\noindent The observed ratio $m_b/m_\tau$ plausibly derives from equality
at the unified scale, as required in all unification schemes extending $SU(5)$.  
The large value of the ``Dirac'' neutrino mass, fixed to $m_t$ through an
$SO(10)$ relation, played a role in our earlier discussion of neutrino masses.

For the lighter quarks the pattern is murkier, as might be expected, since
their masses and mixings are buffeted by subleading effects in the mass
matrices.   However there are some significantly successful attempts to go
further, by exploiting group-theoretic constraints among matrix elements that
arise if one assumes minimalistic gauge symmetry breaking (condensates in
small representations) and simple coupling patterns\cite{ref3}.

\subsection{Emergence of the Planck Scale}
\noindent We can attempt to extend the calculation of Figure~\ref{f9} 
further, to include gravity.  Whereas the couplings of the Theory of Matter
are dimensionless, and run with energy only logarithmically, due to vacuum
polarization effects, the gravitational coupling has dimension (1/mass)$^2$. 
Therefore it grows in importance with energy, even classically -- and much
faster.    A simple-minded estimate, using dimensional analysis, indicates that
the effective gravitational coupling becomes strong at $Q \sim 10^{18}$ Gev,
the Planck mass.  The other couplings unify at $Q \sim 10^{16}$ Gev, and
plausibly become strong at a slightly higher energy.  Thus the unification of
couplings calculation, naively extended to include gravity, is not far off.   This
is quite a remarkable result, since the physical ingredients entering into the
calculation are so disparate.   The small residual discrepancy between the
Theory of Matter unification scale and the Planck scale has been ascribed to
the  opening up of an extra spatial dimension near these scales, though of
course in the present state of knowledge there are other possibilities.  

A dramatic, but I think not unfair, way to state this result is that we have,
within this framework, convincingly solved the central ``hierarchy problem''
of fundamental physics.    By that I mean the question of why gravity, acting
between life-size lumps of matter, is so feeble.  Or, in more technical
language, the problem of why the ratio of the Planck mass to the proton mass
is so large.  In our calculation this ratio is given as the inverse of exponentials
of inverses of the observed coupling constants in the Theory of Matter.  No
spectacularly small (``unnatural'') quantities are involved. The big ratio of
mass scales arises basically because the strong coupling
$\alpha_3$ at the unification scale is about 1/25, and the couplings run only
logarithmically.  Therefore quite a long run is required before one reaches the
scale where $\alpha_3$ approaches unity, protons are assembled, and
ordinary life begins.

Of course other major hierarchy problems (doublet-triplet splitting, smallness
of the soft supersymmetry breaking, $\mu$ problem), more recondite but
still fundamentally significant, remain open, as I've mentioned before.

\section{Is It Right?}
\noindent As I've now discussed, I think we've been given some excellent
clues for figuring out a substantial chunk of physics beyond the Standard
Model.   They point us in the directions of gauge unification and low-energy
supersymmetry. How will we find out whether these ideas or right -- or kill
them off for good?

\subsection{Small Effects Among Known Particles}
\noindent While low-energy supersymmetry features weak couplings and
good ultraviolet behavior, it also introduces a profusion of new particles, with
attendant possibilities for introducing new contributions to flavor-changing
neutral processes, CP violation, and of course diagonal  radiative corrections.

The first class is exemplified by $K-\bar K$ mixing and
$B \rightarrow s\gamma$, to mention two processes that are particularly
sensitive and have received a lot of attention.   We should also include $\mu
\rightarrow e\gamma$ and allied processes in this class.  The second class is
exemplified primarily by electric dipole moments of neutrons or electrons. 
The third class is exemplified primarily by corrections the muon anomalous
magnetic moment
$g_\mu -2$.   The modern experimental limits on deviations from the
Standard Model in each of these processes puts very significant pressure on
the supersymmetric parameter space already.  It is very important to
continue improving these limits.

It may be useful to mention that quantitative interpretation of many
experiments in this field is limited by the accuracy with which we can
calculate even rather simple strong matrix elements.   The technique of lattice
gauge theory, and the available computing power, have markedly improved
recently.   Given the appropriate investments, there could be considerable
progress on this front before long.

\subsection{Proton Decay}
\noindent The Standard Model has the beautiful feature that all baryon- and
lepton-number violating processes require non-renormalizable interactions.
Such interactions are characterized by  coupling constants whose dimensions
are inverse powers of masses.  If the masses involved are extremely large, we
can have a simple universal explanation of the smallness or rarity of such
processes.   Of course, the unification of couplings calculation does suggest
that the relevant mass scale is extremely large.

As was realized very early in the modern history of gauge unification, two
processes above all are exquisitely sensitive to highly suppressed
interactions.  These are neutrino oscillations and proton decay.  Neutrino
oscillations have now been observed, with roughly the predicted oscillation
length (neutrino mass), as I've already discussed.  We're waiting for the other
shoe to drop.

The minimal implementation of gauge unification, without supersymmetry,
has a severe problem with modern experimental limits on proton decay.

Extension of the Standard Model to incorporate low-energy supersymmetry
changes the situation considerably.   The scale of unification goes up a bit,
which removes the outright contradiction between the rate of proton decay
through gauge particle exchange and experiment that we had without
supersymmetry.   On the other hand, dangerous new sources of proton decay
arise, through exchange of the Higgsinos associated with unified symmetry
breaking.   Quantitative analysis is complex and fraught with uncertainties,
but it will not be easy to reconcile limits $\tau_{\rm proton}
\gtrsim 10^{34}$ yrs. with straightforward models.   A striking prediction
characteristic of supersymmetric unified theories is that modes involving
strange final states, particularly $p \rightarrow K^+ \bar\nu$ and
$n \rightarrow K^0 \bar\nu$, will dominate.

\subsection{Focus Point}
\begin{figure}[ht]
\centerline{\BoxedEPSF{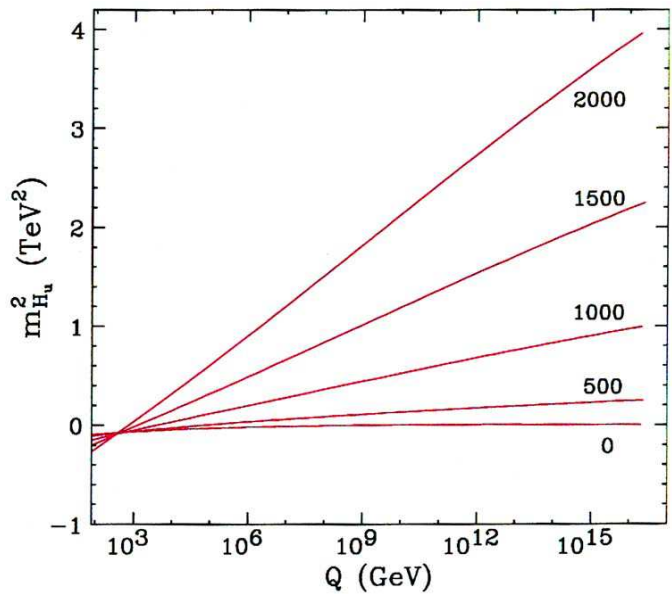 scaled 700}}
\caption{The scale of electoweak symmetry breaking is  surprisingly
insensitive to the value of the scalar mass parameter 
$m_0$.   This is the ``focus point'' phenomenon.  It makes large  values of the
physical squark and slepton masses more plausible.}\label{f12}
\end{figure}

\noindent An interesting recent development, which seems capable of easing
the quantitative pressure on low-energy supersymmetry from all these
sources, is the ``focus point'' scenario of Feng, Matchev, and Moroi\cite{ref4}.  
The central phenomenon is shown in Figure~\ref{f12}.   One finds that the
predicted value of the weak scale is remarkably insensitive to the assumed
value of the soft supersymmetry breaking parameter $m_0$.   Consequently,
that parameter can be taken much larger than one might naively have
expected.

The physical consequence is that squark and slepton -- but not gaugino --
masses can be significantly larger than was previously believed to be
natural.  Masses of 2 Tev are  comfortably allowed.   These larger masses
systematically suppress all the unobserved possibilities mentioned above.

\subsection{Dark Matter Searches}
\begin{figure}[ht]
\centerline{\BoxedEPSF{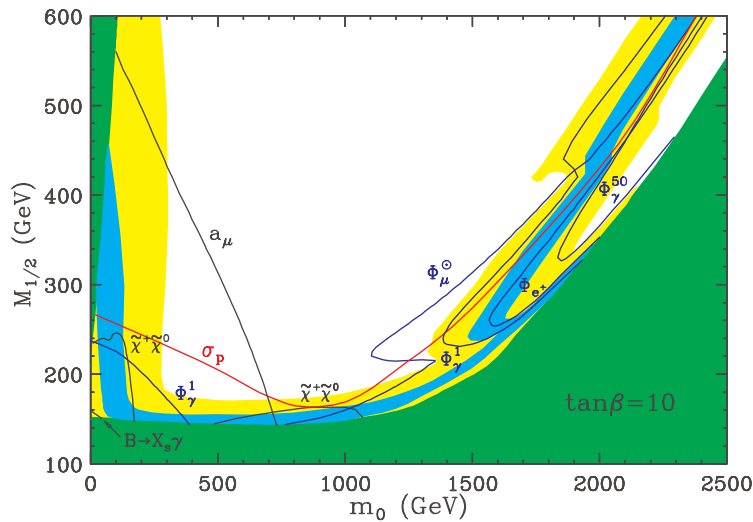 scaled 525}}
\caption{Sensitivity of a number of potential pre-LHC experiments  to
low-energy supersymmetry signatures, in minimal supergravity  models,
allowing for large values of $m_0$.  The blue region  corresponds to
cosmological production of neutralinos with mass  density that could address
the astronomers' missing matter problem.   For detailed explanation, see
Feng,  Matchev, and  Wilczek\protect\cite{fmw}.}\label{f13}
\end{figure}

\begin{figure}[ht]
\centerline{\BoxedEPSF{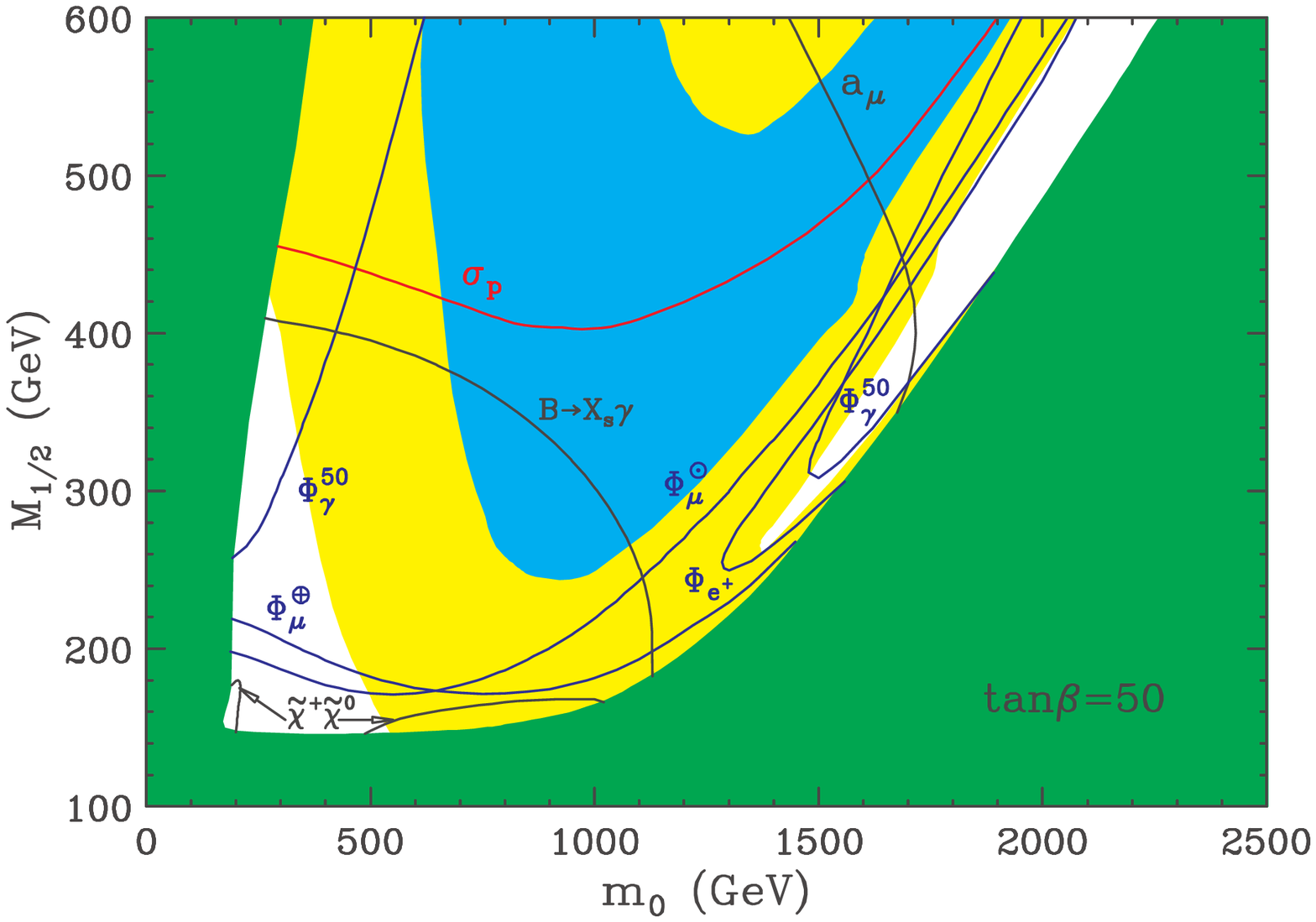 scaled 525}}
\caption{The same as Figure 13, but with the parameter $\tan 
\beta ~=~ 50$ (instead of $\tan \beta ~=~ 10$).}\label{f14}
\end{figure}

\noindent Many of the ideas I have just discussed come together in
Figures~\ref{f13} and
\ref{f14}.  There we see: first, that a wide swath of supersymmetric
parameter space, including a big contribution from the focus point region,
gives rise to a desirable dark matter density; and second, that a wide variety
of experiments for dark matter detection, together with direct accelerator
searches and foreseeable improvements in $B\to s\gamma$ and $g_\mu -2$,
should plausibly give some indication for low-energy supersymmetry even
before the LHC.

\subsection{Produce the New Particles!}
\noindent Of course, the ultimate test for low-energy supersymmetry will be
to produce some of the predicted new $R$-odd particles.  Even in the focus
point scenario, there must be several accessible to the LHC.

\section{Ultimate Questions}
\noindent Finally I'd like briefly to discuss a few questions that definitely
belong on the agenda of future physics, although they fall somewhat outside
the circle of ideas I've been developing so far.

\subsection{Can We Understand Extreme Conditions?}
\noindent We know the equations of QCD, but there are several potentially
awesome applications which await better solutions of those equations.  At
present the theory of neutron star interiors, supernovae explosions, and
leading models of gamma ray bursters are based on crude phenomenological
models of the high-temperature and high-density behavior of hadronic
matter.  It is a great challenge to do better theoretically.   Some quite
beautiful concepts have emerged already, including the liberation of quarks
and gluons in a plasma phase, and the prediction of color superconducting
phases with remarkable properties.  We can test our mettle on experimental
simulation of the Big Bang, in relativistic heavy ion collisions.   Eventually, we
may hope that neutrino and gravity-wave detectors will give us meaningful
access to the most extreme astrophysical processes (a nearby supernova
would be very helpful!).

There has been spectacular progress recently in observational cosmology,
especially in the determination of cosmic microwave background
anisotropies.   These observations seem to indicate the spatial flatness of the
Universe and an approximately scale-invariant fluctuation spectrum, which
broadly supports simple models of inflation.   The general idea of inflation, of
course,  came out of particle physics.   Inflation is supposed to be triggered by
a phase transition associated with breaking of some fundamental symmetry
-- perhaps the unification symmetry.   But existing models of inflation are
only very thinly rooted in specific world-models, and their main parameters
have not been related to microphysics.  It is a great challenge either to show
that inflation really occurs as a consequence of fundamental physics, and to
find an identity and a torso for the inflaton and its potential; or else to replace
it with something different (this is still not inconceivable, perhaps, since the
``evidence'' is rather generic).   Upcoming experiments mapping out the
accurate fluctuation spectrum, and particular polarization measurements
capable of separating out the gravity wave spectrum, will be very interesting
to watch.

The highest energy cosmic rays, including perhaps a neutrino component, will
for the foreseeable future provide us with our highest-energy collisions.  It
would be wonderful to exploit this resource more fully.  Even their origin is
still an open problem, and might involve new fundamental physics.

\subsection{Are the  ``Fundamental" Couplings Universal?}
\noindent According to the basic hypothesis of inflation, the presently
observable Universe arose from the rapid expansion of a tiny patch early on.  
This hypothesis was largely motivated by the ``horizon  problem'': the
observation that observable Universe, as characterized by the large-scale
distribution of galaxies and the cosmic microwave background, is accurately
homogeneous and isotropic.  Inflation guarantees these uniformities, even in
the absence of any dynamics enforcing them, by postulating a common origin
for the observable patches of the sky.   There is no implication that
uniformity characterizes the entire Universe; only the part we presently
observe (with an unknown ``safety factor'' to spare).  

Given this context, it is natural to ask whether other of the observed
uniformities of Nature are likewise cosmologically conditioned.  Specifically,
one can wonder whether quantities we ordinarily regard as ``constants of
Nature'',  such as particle masses and mixing angles, are truly universal, or
instead have frozen-in values that vary from patch to patch.

This phenomenon arises quite concretely in axion physics.  If the
Peccei-Quinn transition occurs before inflation, different amplitudes of the
axion field will be frozen into different patches.   Observers in (very) widely
separated portions of the Universe would report different values of a
fundamental constant of Nature, namely the QCD $\theta$ parameter.
Eventually, as the Universe cooled, the amplitudes would relax, producing
different cosmological mass densities of axions in the different patches. Some
large portions of the Universe would be axion-dominated, while others will
have only a small axion density.

There are many theoretical suggestions for promoting other constants of
Nature into dynamical variables.  It may be attractive, for example,  to
suppose that the symmetry one would have among different families in the
absence of Higgs couplings is only  spontaneously broken.  In that case one
would have axion-like `familons', with similar possibilities to those in the
previous paragraph.   In string theory, it is commonly assumed that all the
physical constants are fields capable of variation.  And many apparently
consistent solutions of the static equations have been found, that predict
wildly different versions of the laws governing observable (low-energy)
physics.

These considerations emphasize the significance of experiments to look for
very light, very weakly interacting particles, the quanta of physical constants
which are actually dynamical variables.   Recently there have been
remarkable improvements in the search for new macroscopic
forces\cite{ref6}.  We look forward to additional experiments looking for
small violations of the equivalence principle and for monopole-dipole forces,
in particular.

\subsection{Why is Empty Space  (Almost) Weightless?}
\noindent The smallness of the cosmological term, compared to other scales of
physics, is notorious.  The vacuum energy density, as seen by gravity, is not
more than
$\sim 10^{-12}$ eV$^4$.  Recent observations suggest it is not zero.  In any
case, it is many orders of magnitude below the Planck or unification energy
density scales
$\sim  10^{108}$ eV$^4$, $\sim 10^{96}$ eV$^4$, the weak energy density
scale $\sim 10^{44}$ eV$^4$, or even the QCD spontaneous chiral symmetry
breaking scale
$\sim 10^{32}$ eV$^4$.

We do not understand the disparity.  In my opinion, it is the biggest and
worst gap in our current understanding of the physical world.

The problem has both classical and quantum aspects.  Classically, it is very
difficult to understand why gravity does not notice the presence of various
symmetry-breaking condensates.  Quantum mechanically, we must also worry
about the energy associated with zero-point oscillations of modes of quantum
fields.   Cancellations due to supersymmetry could partially address the
quantum mechanical aspect, but do not help with the classical aspect; in any
case, supersymmetry in Nature is nowhere near accurate enough for the job.
Even  if we do assume that supersymmetry (or something else) takes care of
the quantum zero-point energy from high-energy modes, we are still left with
the classical contributions, and the contributions of low-energy modes.

Prior to the recent apparent discovery of a non-zero value for the
cosmological term, it was tempting to suppose that some hidden symmetry
somehow put it to zero.   Perhaps now it seems more likely that a dynamical
mechanism is involved.   Some of us\cite{ref7} like the idea that it will involve
relaxation through some exotic, very light, very weakly coupled matter -- in
the spirit of axions relaxing the $\theta$ term -- though I freely admit that
our detailed realization needs work.   In any case, ideas like this reinforce the
interest of searches for new macroscopic forces, and will surely suggest other
sorts of experiments.

The question of why empty space weighs so little is every bit as fundamental
as the question of insuring good ultraviolet behavior of quantum gravity. 
Furthermore, it is much more sharply posed by Nature.   It would be
marvelous if string theory, which promises to provide a unique and consistent
theory of quantum gravity, could meaningfully engage this question.

\section{Future Summary}
\noindent I believe we are about to experience a new Golden Age in
fundamental physics.  The physics of electroweak symmetry breaking,
low-energy supersymmetry, and unification is ripe.  Its fruit will include
Higgs particles, superpartners galore, identification of the dark matter, proton
decay, and more.  The astronomers will chip in with detailed information
about the primordial fluctuations, perhaps including a gravitational wave
component, and perhaps some surprises from high-energy cosmic rays.  As
this tide of discoveries rolls in, we will understand the world better in many
concrete ways.   We will also gather precious information about physics at the
unification scale, and about physical events in the earliest moments of the Big
Bang.

There may also be a Golden Age in the exploitation of the fundamental
physics we have recently achieved.   QCD is a young theory, and not an easy
one to handle.   But continuing advances in computing power, and in
fundamental algorithms (notably, in maintaining chiral symmetry while
discretizing\cite{ref8}), have brought us to the point that definitive
quantitative calculations of a host of quantities are becoming feasible.  For
example a fully microscopic calculation of the proton-neutron mass
difference, so crucial to the structure of the world, would be a milestone
achievement, and is well within sight.  There is a ferment of ideas in
understanding QCD at high temperatures and at high density; it seems
realistic to hope that we will produce usable predictions for the structure of
neutron stars and for behavior in extreme astrophysical environments.

Less easy to anticipate with confidence, but to me a very real and exciting
prospect, is progress on the  frontier of ultra-light, ultra-weakly interacting
matter, with ``firm'' connections to the strong CP problem and somewhat less
firm connections to the vacuum selection and cosmological term problems.

So I expect that in ten to fifteen years we will know a lot more.  Will we know
Everything?  More likely, I think, is that as we learn many additional facts,
we will also come to  comprehend more clearly how much we don't know --
and, let us hope, learn an appropriate humility.

\nonumsection{Acknowledgments}
\noindent This work is supported in part by funds provided by the U.S.
Department of Energy (D.O.E.) under cooperative research agreement
\#DF-FC02-94ER40818.    I would like to thank Jonathan Feng for helpul
comments on the manuscript.

\nonumsection{References}
\noindent These references only touch a few very recent or specialized
topics.  Given the scope of the talk, and the manuscript deadline, this was the
only practical solution for me.  My congratulations to authors of classic papers
too well known to need citation here.

\end{document}